\documentclass[english]{article}
\usepackage{ae,aecompl}
\usepackage[T1]{fontenc}
\usepackage[latin9]{inputenc}
\usepackage{prettyref}
\usepackage{amsmath}
\usepackage{amssymb}

\makeatletter

\usepackage{geometry}

\usepackage{prettyref}
\usepackage[nosort]{cite} 

\usepackage{esint}

\makeatletter

\usepackage{ae}

\usepackage{aecompl}

\usepackage[sf,bf,small]{titlesec}

\DeclareMathOperator{\sech}{sech}

\makeatother

\makeatother

\usepackage{babel}

\begin{document}
\begin{flushright}
{\footnotesize \vspace{30pt}BROWN-HET 1567}\\
{\footnotesize arXiv:0810.4548}
\par\end{flushright}{\footnotesize \par}

\begin{center}
{\LARGE \vspace{15pt}$N$-body Dynamics of Giant Magnons in $\mathbb{R}\times S^{2}$}
\par\end{center}{\LARGE \par}

\begin{center}
{\large \vspace{6pt}Inês Aniceto and Antal Jevicki}
\par\end{center}{\large \par}

\begin{center}
{\large \vspace{3pt}October 26, 2008\vspace{15pt}}
\par\end{center}{\large \par}

\begin{abstract}
We pursue the question of multi-magnon dynamics, focusing on the simplest
case of magnons moving on $\mathbb{R}\times S^{2}$ and working at
the semiclassical level. Through a Pohlmeyer reduction, the problem
reduces to another well known integrable field theory, the sine-Gordon
model, which can be exactly described through an N-body model of Calogero
type. The two theories coincide at the level of equations of motion,
but physical quantities like the energies (of magnons and solitons)
and the associated phase shifts are different. We start from the equivalence
of the two systems at the level of equations of motion and require
that the new (string theory) model reproduces the correct magnon energies
and the phase shift, both of which differ from the soliton case. From
the comparison of energies we suggest a Hamiltonian, and from requiring
the correct phase shift we are led to a nontrivial Poisson structure
representing the magnons.
\end{abstract}

\section{Introduction}

Recent progress in understanding the Gauge/String duality in $N=4$
Super Yang-Mills theory resulted in complete specification of the
worldsheet S-matrix and the associated spectrum \cite{Minahan:2002ve,Beisert:2003ea,Arutyunov:2004vx,Staudacher:2004tk,Beisert:2004hm,Kruczenski:2004kw,Beisert:2005di,Beisert:2005tm,Klose:2006zd,Gromov:2006dh,Beisert:2007ds,Dorey:2007xn,Gromov:2007fn}.
The conjectured exact result received impressive confirmation in both
weak coupling Yang-Mills theory calculations and also semiclassical
string theory calculations at strong coupling \cite{Frolov:2002av,Frolov:2003qc,Frolov:2003tu,Beisert:2003xu,Callan:2003xr,Minahan:2006bd,Hernandez:2006tk,Papathanasiou:2007gd,Chen:2007vs,Gromov:2007aq}.
These successes were accomplished due to the integrability property
characterizing the string dynamics and present in Yang-Mills theory
through its spin-chain representation \cite{Minahan:2002ve,Beisert:2004ry,Beisert:2004ag,Beisert:2005bm,Kazakov:2004qf,Gromov:2007cd}.
At the spectrum level there is a complete classification of states
in terms of magnon excitations. Their dispersion formula is again
known from both weak and strong coupling studies \cite{Beisert:2006qh,Hofman:2006xt}.

Even though all orders results have been accomplished, further study
of the models and of their integrability structures is still desirable.
For instance the spin chain Hamiltonian is reliably known only from
weak coupling calculations, its comparison (and agreement) with the
string theory Hamiltonian is to some degree purely accidental. In
this work we will pursue the question of multi-magnon dynamics. Magnons
scatter from each other with known computable phase shifts \cite{Arutyunov:2004vx,Hernandez:2006tk,Chen:2006gq,Janik:2006dc,Arutyunov:2006iu,Beisert:2006ez,Freyhult:2006vr,Ishizeki:2007kh}
and it is of relevance to determine their interactions. We will do
that in the simplest case of magnons moving on $\mathbb{R}\times S^{2}$
working at the the semiclassical level. The equations of motion in
this case (in a timelike-conformal gauge) coincide with those of the
$O\left(3\right)$ nonlinear sigma model. Multi-magnon solutions have
been constructed in these case using several different techniques,
involving the dressing and the inverse scattering methods \cite{Chen:2006gq,Zakharov:1973pp,Harnad:1983we,Spradlin:2006wk,Kalousios:2006xy}.

One particular approach (Pohlmeyer reduction method) reduces the problem
at the equation of motion level to a well known integrable field theory,
the sine-Gordon model \cite{Pohlmeyer:1975nb}. In this reduction
the role of magnons is played by sine-Gordon solitons. Much is known
about inter-soliton dynamics in the sine-Gordon theory \cite{Faddeev:1977rm}.
In particular it can be exactly described through an N-body model
generalizing the Calogero-Moser model \cite{Calogero:1975ii,Moser:1975qp}.
The relativistic Ruijsenaars-Schneider model \cite{Ruijsenaars:1986vq}
is completely integrable, it summarizes the N-soliton (and anti-soliton)
dynamics for a given coupling and can be directly deduced from sine-Gordon
theory itself \cite{Babelon:1993bx}. In turn it can be used as a
full dynamical theory, even at the quantum level \cite{Ruijsenaars:1986pp}.
It is our goal to establish a related dynamical description for string
theory magnons.

The connection between string dynamics and sine-Gordon theory, is
known to be highly nontrivial. The two theories coincide at the level
of equations of motion, but that is where the comparison stops \cite{Hofman:2006xt}.
Physical quantities like the energies (of magnons and solitons) and
the associated phase shifts are different and it is our intention
to clarify somewhat this nontrivial relationship. The nontrivial dynamical
connection between the two systems can be traced back to a (nonabelian)
dual description of sigma models and the fact that it is in the dual
formulation that the connection can be described in canonical terms.
This was established in several works by Mikhailov\cite{Mikhailov:2005qv,Mikhailov:2005sy,Mikhailov:2005zd}
and remains to be pursued at the quantum level.

For the question of formulating the dynamical system describing multi-magnon
dynamics we start from the fact that at the level of equations of
motion it coincides with the soliton or rather the N-body RS model.
We then require a further fact, namely that the string theory model
ought to reproduce the correct magnon energies and the phase shifts,
both of which differ from the soliton case. From the comparison of
energies we suggest a Hamiltonian, as the $n=-1$ member of the infinite
Hamiltonian sequence \cite{Magri:1978,Ruijsenaars:1989,Oevel:1989}.
Requiring the correct phase shift we are led to a nontrivial Poisson
structure representing the N- magnon dynamics.

The content of the paper is as follows. In \prettyref{sec:Semiclassical-Giant-Magnons}
we give a summary of various classical magnon results in $\mathbb{R}\times S^{2}$
and comparison with sine-Gordon solitons. In \prettyref{sec:Review-of-Sine-Gordon}
we review the integrable dynamics of solitons in terms of the N-particle
R-S description. In \prettyref{sec:An-ansatz} we consider the analogous
representation for magnons. From comparison of energy eigenvalues
we are led to an N-body Hamiltonian given by the inverse of the lax
matrix of the RS model. Elaborating on the phase shift we are led
to suggest a need for an alternative symplectic form. This symplectic
form is explicitly given in the limit of well separated magnons in
\prettyref{sec:Alternative-Poisson-Structure}. \prettyref{sec:Conclusions-&-Acknowledgments}
is reserved for the conclusions.

\section{Semiclassical Giant Magnons in $\mathbb{R}\times S^{2}$\label{sec:Semiclassical-Giant-Magnons}}

String dynamics in the $AdS_{5}\times S^{5}$ space-time can be described
by the $\sigma-$model action\[
S=\frac{\sqrt{\lambda}}{2\pi}\int\mbox{d}\tau\,\mbox{dx}\left\{ \underbrace{\eta^{ab}\partial_{a}Y^{\mu}\partial_{b}Y_{\mu}+\alpha_{1}\left(Y^{2}+1\right)}_{AdS_{5}}+\underbrace{\eta^{ab}\partial_{a}X^{i}\partial_{b}X_{i}+\alpha_{2}\left(X^{2}-1\right)}_{S^{5}}\right\} ,\]
 where one embeds both the sphere as the $AdS$ space in $\mathbb{R}^{6}$
with the respective constraints. This action has a symmetry under
$\mathfrak{su}\left(2,2\right)\times\mathfrak{so}\left(6\right)$.
To consider magnons moving on the sphere one restricts the space-time
to $\mathbb{R}\times S^{5}$, with $\mathbb{R}$ being one of the
time directions of the $AdS_{5}$ space, the respective charges are
the generators of rotations in $S^{5}$ {\small \[
J_{ij}=\frac{\sqrt{\lambda}}{2\pi}\int_{0}^{2\pi}dx\left(X_{i}\dot{X}_{j}-X_{j}\dot{X}_{i}\right),\]
 }and the generator of time translations {\small \[
\Delta=\frac{\sqrt{\lambda}}{2\pi}\int_{0}^{2\pi}dx\dot{Y^{0}}.\]
 }{\small \par}

One considers the limit when the angular momentum $J=J_{12}$ in the
direction $\varphi\equiv(12)$ of $S^{5}$ is very large, and look
at states with $\Delta-J$ finite. The momentum of the excitation
$p$ is also kept fixed.

In this limit, we find that the string ground state has $\Delta-J=0$,
which consists of a point-particle with a light-like trajectory along
the direction $\varphi$, time coordinate $Y^{0}\equiv t$ obeying
$\varphi-t=\mbox{constant}$, and sitting at the origin of the spatial
directions of $AdS_{5}$.

To find excitations above this ground state one looks at solutions
rotating in the $Z_{1}=X^{1}+iX^{2}$ plane. The remaining four directions
of the embedding space we call $\vec{X}$, and $Y^{0}\equiv t$ is
the time co-ordinate (ultimately from $AdS$). So the motion is all
in the time direction of $AdS$ space, and on the subspace $S^{2}\subset S^{5}$:
$\mathbb{R}\times S^{2}$. By choosing a time-like $t=\tau,$ conformal
gauge (the induced metric is proportional to the standard metric,
$\partial_{a}X^{\mu}\partial_{b}X^{\nu}\eta_{\mu\nu}\propto\eta_{ab}$)
we are looking for solutions that solve the Virasoro constraints \[
\left(\partial_{\tau}X^{i}\right)^{2}+\left(\partial_{x}X^{i}\right)^{2}=1,\qquad\partial_{\tau}X^{i}\partial_{x}X^{i}=0,\]
 and obey the conformal equations of motion \[
\left(-\partial_{\tau}^{2}+\partial_{x}^{2}\right)X^{i}+X^{i}\left(-(\partial_{\tau}X^{j})^{2}+(\partial_{x}X^{j})^{2}\right)=0.\]

Solving the equations Hofman and Maldacena \cite{Hofman:2006xt} found
the Giant Magnon solution:\begin{eqnarray*}
Y^{0} & = & \tau,\\
Z_{1} & = & e^{i\tau}\left(c+i\sqrt{1-c^{2}}\tanh u\right),\\
\vec{X} & = & \vec{n}\sqrt{1-c^{2}}\mbox{sech}u,\end{eqnarray*}
 where $c=\cos(p/2)$ is the worldsheet velocity, and $(u,v)$ are
boosted worldsheet co-ordinates\begin{align*}
u & =\gamma(x-c\tau),\\
v & =\gamma(\tau-cx),\qquad\mbox{with }\gamma=\frac{1}{\sqrt{1-c^{2}}}=\frac{1}{\sin(p/2)}.\end{align*}
 This is a rigidly rotating string along the equator of $S^{2}$,
with cusps touching this equator and moving at the speed of light.

In the conformal, time-light gauge we started from, the relevant charges
can be written as \begin{align*}
\Delta & =\frac{\sqrt{\lambda}}{2\pi}\int dx\:1 &  & \mbox{time-translations,}\\
J & =\frac{\sqrt{\lambda}}{2\pi}\int dx\:\mbox{Im}\left(\overline{Z}_{1}\partial_{t}Z_{1}\right) &  & \mbox{angular momentum in }Z_{1}\mbox{ plane,}\\
p & =\frac{1}{i}\int dx\frac{d}{dx}\ln Z_{1} &  & \mbox{worldsheet momentum.}\end{align*}
 For the case of the Giant Magnon, $\Delta$ and $J$ are infinite,
with \begin{equation}
E_{mag}=\Delta-J=\frac{\sqrt{\lambda}}{\pi}\sin(p/2).\label{eq:Energy of GM}\end{equation}

It is well-known that the theory of classical strings moving on $\mathbb{R}\times S^{2}$
is related to the sine-Gordon model at the level of the equations
of motion. For a conformal gauge solution  with $Y^{0}=t$, one has the
Pohlmeyer \cite{Pohlmeyer:1975nb} identification of a scalar field $\alpha(x,t)$as
\begin{align*}
\cos\alpha & =-\partial_{\tau}X^{i}\partial_{\tau}X^{i}+\partial_{x}X^{i}\partial_{x}X^{i}\end{align*}
 obeying the sine-Gordon equation\[
-\partial_{\tau}\partial_{\tau}\alpha+\partial_{x}\partial_{x}\alpha=\sin\alpha\,.\]

The point particle is mapped to the vacuum $\alpha=0$, with zero
energy, while the giant magnon is mapped to the simple kink \cite{Hofman:2006xt}\begin{align*}
\alpha & =4\arctan\left(e^{-\gamma(x-c\tau)}\right),\end{align*}

whose energy is ($\hat{\theta}$ is the asymptotic rapidity of the
sine-Gordon soliton) \begin{equation}
\epsilon_{s.g}=\gamma=\frac{1}{\sin(p/2)}=\cosh\hat{\theta}\,.\label{eq:Energy of sG kink}\end{equation}
 The comparison between sine-Gordon model and the classical string
theory solutions was seen also for another type of {}``dual\char`\"{}
solutions called single spike solutions (see for example \cite{Ishizeki:2007kh,Ishizeki:2007we,Abbott:2008yp}).

But the direct  connection between these two theories holds only  at the level of
equations of motion, and it is non-trivial at the canonical level.
Several physical properties are different, in particular the energies
and the semiclassical phase shifts. In fact the energies shown above
in \prettyref{eq:Energy of GM} and \prettyref{eq:Energy of sG kink}
exibit an inverse relationship

\[
E_{magnon}=\frac{\sqrt{\lambda}}{\pi}\frac{1}{\gamma}=\frac{\sqrt{\lambda}}{\pi}\frac{1}{\epsilon_{s.g}}\,.\]
 This relation can be generalized further for the scattering solution
of two magnons. The two-magnon scattering state, obtained via dressing
method \cite{Spradlin:2006wk} can can be mapped through Pohlmeyer's
reduction to the scattering solutions of two solitons,
whose scattering solution can found in \cite{Faddeev:1977rm}. It
can be seen that the energy of the two-magnon scattering solution
is related to the energy of each of the solitons in the following
way\[
E_{2-mag}=\frac{1}{\epsilon_{s.g,1}}+\frac{1}{\epsilon_{s.g,2}}.\]
 The scattering phase for two magnons is calculated in a very similar
way to the scattering of two sine-Gordon solitons \cite{Jackiw:1975im,Gervais:1976wr}.
See also \cite{Faddeev:1977rm} for a review on the classical and
semiclassical behavior of sine-Gordon solitons. This is not surprising
due to the equivalence of the two classical models through Pohmeyer's
map \cite{Pohlmeyer:1975nb} (see also \cite{Mikhailov:2005qv,Mikhailov:2005zd}).
The time-delay and phase shift of scattering of magnons was also studied
through Bethe Ansatz techniques \cite{Hernandez:2006tk,Arutyunov:2004vx,Chen:2006gq,Arutyunov:2006iu,Freyhult:2006vr}.

Since the string and the sine-Gordon equations share a common time
t, it obviously follows that the time-delay of scattering of giant
magnons (on the string worldsheet) and the time delay for the analogous
scattering problem of solitons (in sine-Gordon theory) is the same\[
\Delta t_{sg}=\Delta t_{mag}=\frac{2}{m\sinh\theta}\ln\tanh\theta=\Delta\tau,\]
 It does not mean however that the scattering phase shifts of the
two problems are the same. In fact they differ, due to the difference
in energies stated above. One has the well known relation, where the
derivative of the phase shift with respect to energy equals the time
delay, in the present case :\[
\frac{\partial\delta_{s.g}}{\partial\epsilon_{s.g}}=\Delta\tau=\frac{\partial\delta_{mag}}{\partial E_{mag}}\quad\Rightarrow\quad\delta_{sg}\ne\delta_{mag}.\]
This will imply that a different interaction is responsible for the
behaviour in the two cases.

\section{Review of sine-Gordon Dynamics\label{sec:Review-of-Sine-Gordon}}

The dynamics of sine-Gordon solitons can be summarized by a  relativistic
$N-$body model due to Ruijsenaars and Schneider. These class of  models \cite{Ruijsenaars:1986vq,Ruijsenaars:1986pp}
represent a relativistic generalization of the Calogero-Moser models
\cite{Moser:1975qp,Calogero:1975ii}. The relation between the field
theoretic system of sine-Gordon solitons and the Lax matrix formulation
of the Ruijsenaars-Schneider model was also thoroughly discussed in
\cite{Babelon:1993bx}. In this section we will review some of the
aspects of this relation and give a summary of the needed notation.
For more details and derivations the reader is directed to the original
references.

For establishing the N-body description of soliton dynamics one starts
with the $N$-soliton solution, written as: \[
e^{-i\phi}=\frac{\det\left(1+A\right)}{\det\left(1-A\right)}.\]
 where $A$ is a $N\times N$ matrix with components \[
A_{ij}=2\frac{\sqrt{\mu_{i}\mu_{j}}}{\mu_{i}+\mu_{j}}\sqrt{X_{i}X_{j}}.\]
 The $\mu_{i}$ are the rapidities, and the $X_{i}=a_{i}e^{2\left(\mu_{i}z_{+}+\mu_{i}^{-1}z_{-}\right)}$
are related to the positions of the soliton through the $a_{i}$.
Here we use light-cone coordinates $z_{\pm}=x\pm t$, and $\partial_{\pm}=\frac{1}{2}\left(\partial_{x}\pm\partial_{t}\right)$.
Note that for a soliton or anti-soliton, $\mu$ is real and $a$ pure
imaginary. The breather solution corresponds to a pair of complex
conjugated rapidities $\left(\mu,\overline{\mu}\right)$ and positions
$\left(a,-\overline{a}\right)$.

The sine-Gordon equation can be described by a Hamiltonian system
with the canonical symplectic form ($\pi$ is the conjugate momentum
to the s.G field $\phi$)\[
\Omega_{sg}=\int\pi\wedge d\phi.\]
 This, by direct substitution can be used to deduce the symplectic
form of the soliton variables $\left(a_{i},\mu_{i}\right)$, which
can be seen to reduce to the usual symplectic form after a change
of variables.

Considering the evolution of the system in terms of the null plane
time $z_{+}=\tau$ one has \[
A\left(\sigma,\tau\right)=e^{\sigma\mu^{-1}}\tilde{A}\left(\tau\right)e^{\sigma\mu^{-1}},\]
 where $\left[\mu\right]_{ij}=\mu_{ij}=\mu_{i}\delta_{ij}$ is the
matrix of rapidities, $\tilde{X}_{i}=a_{i}e^{\mu_{i}\tau}$ are the
soliton coordinates. The matrix (coordinate) \[
\tilde{A}\left(\tau\right)=2\frac{\sqrt{\mu_{i}\mu_{j}}}{\mu_{i}+\mu_{j}}\sqrt{\tilde{X}_{i}\tilde{X}_{j}}.\]

is then used to reconstruct the Lax matrix of the N-body system. Through
diagonalization one has:\begin{eqnarray*}
Q & = & U^{-1}\tilde{A}U,\\
L & = & U^{-1}\mu U.\end{eqnarray*}
 where $Q=\mbox{diag}\left(Q_{1},\cdots,Q_{N}\right)$ are the eigenvalues
of $\tilde{A}$. The $N$-soliton solution is then written as $e^{-i\phi}=\prod_{i=1}^{N}\frac{1+Q_{i}}{1-Q_{i}}$,
and the matrix $L$ is the Lax operator, as its time evolution is
given by a (Lax) equation\[
\dot{L}\equiv\frac{dL}{d\tau}=\left[M,L\right],\quad M=\dot{U}U^{-1}.\]
 Consequently, the quantities $H_{n}=\mbox{Tr}\left(L^{n}\right)=\sum_{i=1}^{N}\mu_{i}^{n}$
are conserved through the evolution of solitons.

Finally, if we define $\rho_{i}=\dot{Q}_{i}/Q_{i}$, and perform the
change of variables $\left(\mu_{i},a_{i}\right)\rightarrow\left(Q_{i},\rho_{i}\right)$,
which is a symplectic transformation, we find the Lax matrix to have
the same form as the original $\tilde{A}$: \begin{equation}
L=2\frac{\sqrt{Q_{i}Q_{j}}}{Q_{i}+Q_{j}}\sqrt{\rho_{i}\rho_{j}}.\end{equation}

The Poisson brackets of these two variables $Q_{i},\rho_{i}$ are
not canonical, so it is convenient to introduce a new set of variables
$\theta_{i}$ conjugated to the variables $q_{i}$, given by\[
\rho_{i}=e^{\theta_{i}}\prod_{k\ne i}\frac{Q_{i}+Q_{k}}{Q_{k}-Q_{i}},\; Q_{j}=ie^{iq_{j}}\,.\]
 In these new variables, the original symplectic form $\int\pi\wedge d\phi$
is simply given by the usual $\int\theta_{i}dq_{i}$, which corresponds
to the canonical Poisson brackets. From the sequence of conserved
quantities, or Hamiltonians, $H_{n}=\mbox{Tr}\left(L^{n}\right)$
we are interested in particular in the $H_{\pm1}$, which are the
generators of the evolution in the light cone coordinates $\tau=z_{+}$
and $\sigma=z_{-}$. These are given by \[
H_{\pm1}=\mbox{Tr}\left(L^{\pm1}\right)=\sum_{j}e^{\pm\theta_{j}}\prod_{k\ne j}\left|\coth\left(\frac{q_{j}-q_{k}}{2}\right)\right|.\]
 The full Hamiltonian is given by $H=\frac{1}{2}\left(H_{+1}+H_{-1}\right)=\mbox{Tr}\mathcal{L}_{rs}$,
where we define\begin{equation}
\mathcal{L}_{rs}=\frac{1}{2}\left(L+L^{-1}\right).\label{eq:Hamilt-matrix-Lrs}\end{equation}

This system corresponds to a particular case of the $N-$particle
relativistic Ruijsenaars-Schneider model \cite{Ruijsenaars:1986vq}.
The Lax matrix for the general case of RS model was constructed in
\cite{Ruijsenaars:1986vq}. This Lax matrix is defined by\begin{equation}
L_{j}=V_{i}C_{ij}V_{j},\label{eq:Lax matrix for RS model}\end{equation}
 where\[
V_{i}\equiv e^{\frac{1}{2}\theta_{i}}\left(\prod_{k\ne i}f\left(q_{i}-q_{k}\right)\right)^{1/2},\]
 and $C_{ij}\left(q\right)$ is directly related to the choice of
$f\left(q\right)$. For a family of interaction potentials of the
type given below\begin{equation}
f\left(q\right)=\left[1+\alpha/\sinh^{2}\left(\frac{\mu q}{2}\right)\right]^{1/2}\,,\quad\mu,\alpha\in\left(0,\infty\right),\label{eq:RS function f(q)}\end{equation}
 the components $C_{ij}$ are just given by $C_{ij}\left(q\right)=\left[\cosh\left(\frac{\mu q}{2}\right)+ia\sinh\left(\frac{\mu q}{2}\right)\right]^{-1},$
with $\left(1+a^{2}\right)^{-1}=\alpha^{2}$.

The model has an infinite set of commuting conserved charges \[
H_{n}=\mbox{Tr}\left(L^{n}\right)\,,\; n\in\mathbb{N},\]

with the Hamiltonian (generator of time translations) and momentum
(generator of space translations) given as \cite{Ruijsenaars:1986vq}
\begin{eqnarray}
H & = & \frac{1}{2}\left(H_{1}+H_{-1}\right)=mc^{2}\sum_{j=1}^{N}\cosh\theta_{j}\prod_{k\ne j}f\left(q_{k}-q_{j}\right),\label{eq:RS _Hamilt _lax_mat}\\
P & = & \frac{1}{2}\left(H_{1}-H_{-1}\right)=mc\sum_{j=1}^{N}\sinh\theta_{j}\prod_{k\ne j}f\left(q_{k}-q_{j}\right),\label{eq:RS_mom_lax_mat}\end{eqnarray}
 where $q_{i}$ are the positions of the solitons and $\theta_{i}$
the conjugate rapidities. The interaction between solitons is given
by the even function $f\left(q_{k}-q_{j}\right)$, reducing to $f\equiv1$
in the free theory. The RS model is relativistic, as the generators
($\mathcal{B}$ is the generator for Boosts) \[
\mathcal{H}_{k}=\frac{1}{2}\left(H_{k}+H_{-k}\right),\quad\mathcal{P}_{k}=\frac{1}{2}\left(H_{k}-H_{-k}\right),\quad\mathcal{B}=-\frac{1}{c}\sum q_{j},\]
 obey the two-dimensional Poincaré algebra:\begin{equation}
\left\{ \mathcal{H}_{k},\mathcal{P}_{k}\right\} =0\,,\quad\left\{ \mathcal{H}_{k},\mathcal{B}\right\} =\mathcal{P}_{k}\,,\quad\left\{ \mathcal{P}_{k},\mathcal{B}\right\} =\mathcal{H}_{k}\,.\label{eq:Poincare algebra}\end{equation}

 Next  let us  discuss the question
of the time delay (and phase shift) in  the particle picture. Considering the two particle case, one goes to the
center-of-mass frame, as in \cite{Ruijsenaars:1986vq}:\begin{eqnarray}
s\equiv q_{1}+q_{2} & ,\quad & \varphi\equiv\frac{1}{2}\left(\theta_{1}+\theta_{2}\right),\nonumber \\
q\equiv q_{1}-q_{2} & ,\quad & \theta\equiv\frac{1}{2}\left(\theta_{1}-\theta_{2}\right),\label{eq:N=2 COM coords}\end{eqnarray}

The Lax matrix \eqref{eq:Lax matrix for RS model} and its inverse
are then given by\[
L=e^{\varphi}f\left(q\right)\left(\begin{array}{cc}
e^{\theta} & C_{12}\\
\bar{C}_{12} & e^{-\theta}\end{array}\right)\quad;\qquad L^{-1}=e^{-\varphi}f\left(q\right)\left(\begin{array}{cc}
e^{-\theta} & -C_{12}\\
-\bar{C}_{12} & e^{\theta}\end{array}\right),\]
 where $C_{ij}$ is a $2\times2$ matrix with entries $C_{11}=C_{22}=1$,
and $C_{12}=\bar{C}_{21}=\left[\cosh\left(\frac{\mu}{2}q\right)+ia\sinh\left(\frac{\mu}{2}q\right)\right]^{-1}$.
Now it is simple to check that the Hamiltonian \eqref{eq:RS _Hamilt _lax_mat}
becomes\begin{equation}
H=2\cosh\varphi\cosh\theta\, f\left(q\right)=\left(\cosh\theta_{1}+\cosh\theta_{2}\right)f\left(q\right).\label{eq:Lax_Hamilt_2part}\end{equation}
 The momentum given by \eqref{eq:RS_mom_lax_mat} also becomes:\begin{equation}
P=2\sinh\varphi\cosh\theta\, f\left(q\right)=\left(\sinh\theta_{1}+\sinh\theta_{2}\right)f\left(q\right).\label{eq:Lax _Mom_2part}\end{equation}

One comment should be made with respect to the interaction potential
$f\left(q\right)=\left[1+\alpha/\sinh^{2}\left(\frac{\mu q}{2}\right)\right]^{1/2}$.
Going back to \eqref{eq:RS function f(q)} one can see that for $\alpha=1$
it reduces to the particular case of a repulsive soliton-soliton interaction
in the sine-Gordon model, $f_{r}\left(q\right)=\left|\coth\left(\frac{q}{2}\right)\right|$.
An extension of the interaction potential to $\alpha=-1$ leads to
the attractive case of soliton-anti-soliton interaction of sine-Gordon,
where $f_{a}\left(q\right)=\left|\tanh\left(\frac{q}{2}\right)\right|$.

With the Hamiltonian \eqref{eq:Lax_Hamilt_2part}
and choosing certain interacting potentials $f$ one can fully  recover the
properties of the sine-Gordon soliton- (anti)soliton scattering, such
as time delay and phase shift. From
\eqref{eq:Lax _Mom_2part} it is easy to see that the center of mass
$P=0$ corresponds to $\varphi=0$. Then the center-of-mass Hamiltonian
for two particles is:\[
H_{cm}=\cosh\theta\, f\left(q\right).\]

Now we have the relation:\[
\dot{q}^{2}+f^{2}\left(q\right)=H_{cm}^{2}.\]
 Because $H_{cm}$ is a constant of motion, so is the quantity $\epsilon\equiv H^{2}-1$.
Evaluating $\epsilon$ asymptotically, when $x\to\infty$, we obtain
$\epsilon=\sinh^{2}\hat{\theta}$, where $\hat{\theta}$ is the asymptotic
center-of-mass rapidity. The time delay is then determined by the
time taken along a trajectory from $-q$ to $q$, as $\left|q\right|\to\infty$.
For the repulsive soliton-soliton case $f_{r}$, we get\[
\int_{-q}^{q}\frac{dq}{\sqrt{\epsilon-\mbox{csch}^{2}\left(\frac{q}{2}\right)}}=\left.\frac{4}{\sqrt{\epsilon}}\cosh^{-1}\left(\sqrt{\frac{\epsilon}{\epsilon+1}}\cosh\left(\frac{q}{2}\right)\right)\right|_{2\cosh^{-1}\sqrt{\frac{1+\epsilon}{\epsilon}}}^{q}\:\underset{q\to\infty}{\to}\frac{2q}{\sinh\hat{\theta}}+\frac{1}{\sinh\hat{\theta}}\ln\left(\tanh\hat{\theta}\right).\]
 The first term is the time for each of the solitons to go from $-q$
to $q$ if if it was free (no interaction). The second term is in
fact the time delay due to having a repulsive interaction, and correctly
reproduces the time delay for a soliton-soliton scattering in sine-Gordon
theory, obtained through field theoretic methods.

\section{An ansatz for the dynamics of a two-magnon system\label{sec:An-ansatz}}

Our aim is to describes the N-magnon dynamics in Hamiltonian terms.
The appropriate dynamical system ought to be such that it reproduces
the classical equations of motions, its energy, momentum and finally
phase shift in agreement with the known magnon results \cite{Hofman:2006xt}.
We will begin by focusing on the two-magnon interactions.

We know that the sine-Gordon and the magnons have the same classical
equations of motion, and as such the time delay for both systems agrees
\[
\Delta t_{m}\left(E_{m}\right)=\left.\Delta t_{sg}\left(\epsilon_{sg}\right)\right|_{\epsilon_{sG}=\frac{1}{E_{m}}}\]
 but with different energies. This implies different Hamiltonians
for the two systems. With the semiclassical phase shift obeying $\frac{\partial\delta\left(E\right)}{\partial E}=\Delta t$
one can try to deduce the (Hamiltonian) dynamics directly from the
phase shift itself.

For the sine-Gordon system the center-of-mass Hamiltonian is $H_{sg}=\cosh\theta\, f\left(q\right)=\epsilon_{sg}$,
the equation of motion for the relative position $q$ gives $\dot{q}=\sqrt{\epsilon_{sg}^{2}-f\left(q\right)^{2}}$,
and the time delay in terms of the energy is simply given by\[
\Delta t_{sg}=\int\frac{dq}{\dot{q}}=\int\frac{dq}{\sqrt{\epsilon^{2}-f\left(q\right)^{2}}}\,.\]
 and the scattering phase shift of two sine-Gordon solitons
is just given by $\delta_{sg}\left(\epsilon\right)=\int d\epsilon\,\Delta t_{sg}$,
while for the two-magnons \begin{equation}
\delta_{m}\left(E_{m}\right)=\int dE_{m}\left.\Delta t_{sg}\left(\epsilon_{sg}\right)\right|_{\epsilon_{sg}=\frac{1}{E_{m}}}=\int dE_{m}\, E_{m}\int\frac{dq}{f\left(q\right)\sqrt{f^{-2}\left(q\right)-E_{m}^{2}}}\,.\end{equation}

In order to determine which Hamiltonian $H_{cm}\equiv E_{m}$ produces
this phase shift we first perform a change of variables, introducing
a new coordinate $Q$ through $dQ=\frac{dq}{f\left(q\right)}$. Also
define $F\left(Q\right)=\frac{1}{f\left(q\left(Q\right)\right)}.$
The interaction then follows (soliton-soliton interaction): for f$\left(q\right)=\coth q$
we find $q=\cosh^{-1}\left(e^{Q}\right)$ and $F\left(Q\right)=\sqrt{1-e^{-2Q}}$.

This means that the limit of relative position $q\rightarrow\infty$
corresponds to the new relative position doing the same $Q\rightarrow\infty$.
Also in this limit, we have $f\left(q\right),F\left(Q\right)\rightarrow1$
(the free theory limit).

After this change of variables, we rewrite the phase shift as\[
\delta\left(E_{m}\right)=\int dE_{m}\int dQ\sqrt{\frac{E_{m}^{2}}{F^{2}-E_{m}^{2}}}=\int dE_{m}\int\frac{dQ}{\dot{Q}},\]
 and want to find the center-of-mass Hamiltonian $H_{cm}\equiv E_{m}$
such that\begin{equation}
\dot{Q}=\frac{\partial H_{cm}}{\partial\alpha}=\sqrt{\frac{F^{2}-H_{cm}^{2}}{H_{cm}^{2}}}\,,\end{equation}
 where $\alpha$ is the new relative rapidity, i.e. the conjugate
variable to $Q$. The differential equation above can be solved to
give \[
\sqrt{1-\left(\frac{H_{cm}}{F}\right)^{2}}=-\frac{\alpha}{F}.\]
 This result is only valid for $\alpha<0$. Squaring this result,
we can solve for $H_{cm}$, and find \begin{equation}
H_{cm}=\sqrt{1-\alpha^{2}-e^{-2Q}}.\end{equation}

This two-body magnon Hamiltonian appears to be of relativistic (Toda)
type. It faithfully reproduces the magnon scattering phase shift.
But it is not directly recognizable as a known integrable system.
Furthermore it is not obvious how to extend it to the N-body case.
First one would need to find a two-body Hamiltonian that reduces to
$H_{cm}$ in the center-of-mass. In the limit $Q\rightarrow\infty$
we have that \[
H=\varepsilon_{1}+\varepsilon_{2}=\sin\frac{p_{1}}{2}+\sin\frac{p_{2}}{2},\]
 where $\varepsilon_{1,2}=\sin\frac{p_{1,2}}{2}$ are the energies
of each magnon in the free theory. For only one magnon the Hamiltonian
would be given by $H=\sqrt{1-\alpha^{2}}=\sin\frac{p}{2}$, which
means that the relation between the rapidity $\alpha<0$ and the momentum
$p$ is $\alpha=-\cos\frac{p}{2}=\cos(\pi+\frac{p}{2})$. These results
will hold in the free theory limit for each magnon. Then a good ansatz
for the two-body Hamiltonian, which reproduces the correct result
for the free limit, would be\begin{equation}
H_{2}=\sqrt{1-\alpha_{1}^{2}-e^{-2Q}}+\sqrt{1-\alpha_{2}^{2}-e^{-2Q}},\end{equation}
 with $Q=Q_{1}-Q_{2}$, and the momentum of each magnon is given by
$\frac{p_{i}}{2}=\arccos(\alpha_{i})-\pi$.

This construction is non-unique because we do not have the expression
of the total momentum. As mentioned we also have no information on
the integrability properties of this system, which is fulcral to generalise
our results to the dynamics of $N$-magnon solutions. For these reasons
we now pursue a different strategy, based on employing the known integrable
structure of the RS model, in particular its Lax matrix L. Together
with the classical equivalence between sine-Gordon solitons and giant
magnons there was evidence that the poles of the S-matrix of scattering
magnons were related to a Calogero type system in the non-relativistic
limit \cite{Dorey:2007xn}, thus making us believe that the dynamics
of magnons are intimately related to the dynamics of solitons in the
RS model. In fact one would hope to describe the dynamics of magnons
through a Lax pair formulation whose Lax matrix would be directly
related to the Lax matrix of the relativistic RS model.

\subsection{The $N$-magnon Hamiltonian}
With the motivation for using the RS integrable structure described above
we now proceed to the construction of the associated magnon dynamical system.
This will involve specifying both the hamiltonian and the symplectic structure.
As we have emphasized  before, the energies of the sine-Gordon solitons and the
magnons are inverse of each other. This result leads us to the following
ansatz for the $N$-magnon Hamiltonian:\begin{equation}
H_{m}=\mbox{Tr}\left[\mathcal{L}_{rs}^{-1}\right],\label{eq:Ansatz for 2-mag Hamilt}\end{equation}
 where $\mathcal{L}_{rs}$ is related to the Lax matrix of the RS
model through \prettyref{eq:Hamilt-matrix-Lrs}. We will now study this hamiltonian
and consider the two-magnon interaction.

Recall that from \prettyref{eq:Hamilt-matrix-Lrs} \[
\mathcal{L}_{rs}=\frac{L+L^{-1}}{2}=\frac{f\left(q\right)}{2}\left\{ e^{\varphi}\left(\begin{array}{cc}
e^{\theta} & C_{12}\\
\bar{C}_{12} & e^{-\theta}\end{array}\right)+e^{-\varphi}\left(\begin{array}{cc}
e^{-\theta} & -C_{12}\\
-\bar{C}_{12} & e^{\theta}\end{array}\right)\right\} .\]
 The RS Hamiltonian \prettyref{eq:RS _Hamilt _lax_mat} is just the
trace of the matrix above. This matrix has the following eigenvalues:%
\footnote{For a $2\times2$ matrix $M=\left(\begin{array}{cc}
a & b\\
c & d\end{array}\right),$ its eigenvalues are simply given by $\lambda_{\pm}=\frac{a+d}{2}\pm\frac{1}{2}\sqrt{\left(a-d\right)^{2}+4bc}.$%
}\[
h_{\pm}=\frac{f\left(q\right)}{2}\left(\cosh\left(\varphi+\theta\right)+\cosh\left(\varphi-\theta\right)\pm\sqrt{\left(\cosh\left(\varphi+\theta\right)-\cosh\left(\varphi-\theta\right)\right)^{2}+4\sinh^{2}\varphi\left(1-f\left(q\right)^{-2}\right)}\right).\]
 Then the Hamiltonian for the 2-magnon problem \prettyref{eq:Ansatz for 2-mag Hamilt}
will be just\begin{equation*}
H_{m} \, =\, h_{+}^{-1}+h_{-}^{-1}\, =\,
\frac{1}{f\left(q\right)}\frac{2\cosh\theta\cosh\varphi}{\cosh^{2}\theta+f\left(q\right)^{-2}\sinh^{2}\varphi}.\end{equation*}

Recall that if $M$ is a diagonalizable matrix with $\Lambda=\mbox{diag}\left(\lambda_{1},\cdots,\lambda_{N}\right)$ the
diagonal matrix of eigenvalues, then for a smooth function $g\left(M\right)$
the trace of $g\left(M\right)$, it will be given by\begin{equation}
\mbox{Tr}\left[g\left(M\right)\right]=\mbox{Tr}\left[g\left(\Lambda\right)\right]=\sum_{i=1}^{N}g\left(\lambda_{i}\right).\label{eq:Traces_of_matrices}\end{equation}

It is easy to check that in the free theory ($f\left(q\right)\equiv1$)
we have:\[
H_{m}^{free}=\frac{1}{\cosh\theta_{1}}+\frac{1}{\cosh\theta_{2}}=E_{mag,1}+E_{mag,2},\]
 which corresponds to the sum of the energy of the two magnons, as
expected.

To have an ansatz for the momentum of the $N$-body magnon problem,
we first look at the momentum for the magnons. In \cite{Hofman:2006xt}
we have that for one magnon the relation between the momenta $p_{m}$
and the rapidity $\theta$ is given by $\cosh\theta=\left[\sin\frac{p_{m}}{2}\right]^{-1}.$
But we know that for the sine-Gordon model the total momentum is $P=\sum_{i}p_{i}=\sum_{i}\sinh\theta_{i}.$
Then a simple comparison allows us to conclude that the momenta for
each magnon $p_{m,i}$ is related to the momenta of each soliton $p_{i}$
by:\begin{equation}
\sin\left(\frac{p_{m,i}}{2}\right)=\frac{1}{\sqrt{1+p_{i}^{2}}}.\end{equation}
 Thus, a good ansatz for the momenta of the magnon $P_{m}=\sum_{i}p_{m,i}$
will be \begin{equation}
P_{m}=2\,\mbox{Tr}\left[\arcsin\left(\mathbf{1}+\mathcal{P}_{rs}^{2}\right)^{-1/2}\right],\end{equation}
 where we defined the momentum matrix for the RS model (whose trace
gives the RS momenta given by \eqref{eq:RS_mom_lax_mat}) to be $\mathcal{P}_{rs}=\frac{L-L^{-1}}{2}.$
By knowing the eigenvalues of this last matrix, we can determine $P_{m}$
by using the result \eqref{eq:Traces_of_matrices}:\begin{equation}
P_{m}=2\sum_{i=\pm}\arcsin\left(\frac{1}{\sqrt{1+p_{i}^{2}}}\right).\end{equation}
 The eigenvalues of $\mathcal{P}_{rs}$ are \[
p_{\pm}=\frac{f\left(q\right)}{2}\left\{ \sinh\theta_{1}+\sinh\theta_{2}\pm\sqrt{\left(\sinh\theta_{1}-\sinh\theta_{2}\right)^{2}+4\cosh^{2}\varphi\left(1-f\left(q\right)^{-2}\right)}\right\} .\]

The magnon momentum will then be given by\begin{eqnarray*}
\sin\frac{P_{m}}{2} & = & \frac{p_{+}+p_{-}}{\sqrt{1+p_{+}^{2}}\sqrt{1+p_{-}^{2}}}\,.\end{eqnarray*}

In the limit of the free theory $f\left(q\right)\to1$, we find the
expected relation $P_{m}^{free}=p_{1}^{m}+p_{2}^{m}$, i.e. the magnon
momentum is the sum of the momenta for each magnon.

Note that because all integrals of motion $H_{k}$ Poisson commute
with each other, so will $H_{m}$ and $P_{m}$:\[
\left\{ H_{m},P_{m}\right\} =0.\]

The center of mass condition is given by \[
P_{m}=0\;\Rightarrow\quad\sin\frac{P_{m}}{2}=0\;\Rightarrow\quad\theta_{1}+\theta_{2}=0.\]
 In the center of mass, the Hamiltonian is simply \begin{eqnarray*}
H_{m} & = & \frac{1}{f\left(q\right)}\left(\sech\theta_{1}+\sech\theta_{2}\right)=\frac{2}{f\left(q\right)}\sech\theta,\end{eqnarray*}
 with $f\left(q\right)$ the same as before.

A check on our ansatz is determining the classical and semiclassical
behaviors of our system, such as the time delay and phase shift for
this two-body problem of scattering magnons, and compare them to the
known results \cite{Hofman:2006xt}.

We start from the center-of-mass Hamiltonian determined above, and
determine the classical equations of motion and time delay. But to
do so, we need to choose a Poisson structure. Let us assume that the
Poisson structure is the symplectic one. Then the equation of motion
for $q$ is just

\[
\dot{q}\equiv\frac{\partial H}{\partial\theta}=-H_{m}\,\tanh\theta=H_{m}\sqrt{1-\frac{1}{4}f\left(q\right)^{2}H_{m}^{2}}.\]
 The Hamiltonian is a conserved quantity, $H_{m}\equiv E$ and can
be evaluated when $q\rightarrow\infty$, giving $E=\sech\hat{\theta}$,
where $\hat{\theta}$ is the asymptotic rapidity. We find the time
delay in this case to be \[
\Delta T_{m}=\int\frac{dq}{\dot{q}}=\cosh^{2}\hat{\theta}\Delta T_{RS}.\]
 But this time delay is not correct, nor does it reproduce the right
phase shift.

The phase shift is determined by WKB semiclassical methods to be given
by the symplectic structure $\omega$ (the inverse of the Poisson
structure). In phase space variables $\left(x_{i},p_{i}\right)$ we
have:\begin{equation}
\delta\left(E\right)\equiv\int p_{i}\,\omega_{ij}dx_{j}.\label{eq:phase shift from SF}\end{equation}
 For the results in this section we have used a canonical Poisson
brackets (the standard symplectic structure $\left\{ p_{i},x_{j}\right\} =\delta_{ij}$),
so the phase shift is simply $\delta\left(E\right)=\int\theta\, dq$.
By solving $H_{m}\equiv\frac{2}{f\left(q\right)}\sech\theta=E$, with
respect to the rapidity, $\theta=\cosh^{-1}\left(\frac{2}{f\left(q\right)E}\right)$,
we can determine the semiclassical phase shift to be \[
\delta\left(E\right)=\int\cosh^{-1}\left(\frac{2}{f\left(q\right)E}\right)dq.=\int dE\,\Delta T_{m}.\]

We find that even though our ansatz correctly reproduces the energies
and momenta of the magnon system, it does not give the expected classical
behaviour (time delay or equations of motion) nor the semiclassical
phase shift. But in these calculations we have assumed the usual canonical
Poisson brackets, which is equivalent to having a canonical symplectic
form for $q$ and $p$ and which resulted in the usual form of the
Hamilton-Jacobi equations, namely $\dot{q}=\frac{\partial H}{\partial p}$
and $\dot{p}=-\frac{\partial H}{\partial p}$. One can trace the difference
in phase shifts to the different Poisson structures in the two cases.
The semiclassical phase shift can be related to the symplectic form
$\omega$ (the inverse of the Poisson structure) as follows :\[
\delta=\int p_{i}\omega_{ij}dq_{j}=\int p_{i}\omega_{ij}\dot{q}_{j}dt\,.\]
 For the trivial symplectic structure, $\omega\equiv\mbox{Id}$, and
the phase shift has the usual form. But a non-trivial symplectic form
is required for reproducing the correct phase shifts and for defining
the full magnon N-body dynamics. This we will identify in the next section.

\section{Poisson Structure for the $N$-Magnon Dynamics\label{sec:Alternative-Poisson-Structure}}

 We have identified above the N-body Hamiltonian
for magnons as one member of the RS hierarchy. We have also understood
that a new modified Poisson (and symplectic) structure is needed in
order to obtain both the correct equations of motion and the correct
magnon phase shift. In the present section we will be able to specify
the modified symplectic structure in an approximation of well separated
magnons. This approximation which we take just for the purpose of
simplifying the problem involves a limit of the RS model when the
solitons are far away from each other, called the Toda lattice. The
relativistic Toda lattice was introduced in \cite{Ruijsenaars:1989}
as a relativistic version of the regular Toda lattice \cite{Toda:1981,Olshanetski:1981,Olshanetski:1983}.
In these models the study of master symmetries \cite{Fuchssteiner:1983,Oevel:1987bk,Damianou:1990}
and of recursion relations \cite{Das:1988wm,Bruschi:1988,Oevel:1989}
led to the discovery of a sequence of Hamiltonian/Poisson structures
that return the same classical equations of motion, result of the
existence of a bi-Hamiltonian system \cite{Magri:1978}.

As seen in \cite{Ruijsenaars:1989} we obtain the simpler model of
relativistic Toda Lattice from the original Ruijsenaars-Schneider
model \eqref{eq:RS _Hamilt _lax_mat} by considering that the particles
are very far from each other $q_{i-1}\ll q_{i}$.%
\footnote{In fact Ruijsenaars introduced the limit $\varepsilon\rightarrow0$
of the variables \[
q_{j}^{\varepsilon}\rightarrow q_{j}-2j\ln\varepsilon\:,\quad j=1,\cdots,N.\]
} This allows us to keep only the nearest neighbour interactions and
these interactions become exponential $f\left(q\right)=\sqrt{1+g^{2}e^{q}}$.
Note that we are studying the nonperiodic Toda lattice, for which
$q_{0}=-\infty$ and $q_{N+1}=\infty$.

The Hamiltonian for the relativistic Toda lattice is given by \begin{equation}
H=\sum_{i=1}^{N}e^{\theta_{i}}V_{i}\left(q_{1},...,q_{N}\right).\label{eq:Toda Hamiltonian}\end{equation}
 But now the interaction potential is given by nearest neighbour interactions
only\begin{equation}
V_{i}\left(q_{1},...,q_{N}\right)=f\left(q_{i-1}-q_{i}\right)f\left(q_{i}-q_{i+1}\right),\quad i=1,...N.\end{equation}
 Also, the symplectic form remains \[
\omega=\sum_{i=1}^{N}dq_{i}\wedge d\theta_{i}.\]

This system is integrable and has a Lax matrix formulation, inherited
from the RS model (up to some similarity transformation) \cite{Ruijsenaars:1989,Bruschi:1988,Oevel:1989,Suris:1993,Damianou:1994}.
To write the Lax matrix we introduce the following change of variables
\[
a_{j}=g^{2}e^{q_{j}-q_{j+1}+\theta_{j}}\frac{f(q_{j-1}-q_{j})}{f(q_{j}-q_{j+1})}\quad;\quad b_{j}=e^{\theta_{j}}\frac{f(q_{j-1}-q_{j})}{f(q_{j}-q_{j+1})}\,,\quad j=1,...,N.\]
 Note that $a_{0}=a_{N}=0$. The Lax matrix is then given by\[
L=\left(\begin{array}{ccccc}
a_{1}+b_{1} & a_{1}\\
a_{2}+b_{2} & a_{2}+b_{2} & a_{2} & 0\\
\vdots & \vdots & \ddots & \ddots\\
a_{N-1}+b_{N-1} & a_{N-1}+b_{N-1} & \cdots & a_{N-1}+b_{N-1} & a_{N-1}\\
b_{N} & b_{N} & \cdots & b_{N} & b_{N}\end{array}\right).\]

The Hamiltonian $H_{1}\left(q,p\right)$ given in \eqref{eq:Toda Hamiltonian}
can be written in the new variables:\begin{equation}
h_{1}=\mbox{Tr}L=\sum_{i=1}^{N-1}a_{i}+\sum_{i=1}^{N}b_{i},\label{eq:Linear Hamiltonian}\end{equation}

The equations of motion in the $\left(q,\theta\right)$ coordinates
are given by \[
\dot{q}_{j}=e^{\theta_{j}}V_{j}\quad;\quad\dot{\theta}_{j}=-\sum_{k}e^{\theta_{k}}\frac{\partial V_{k}}{\partial q_{j}},\]
 which can be obtained from the Hamiltonian \eqref{eq:Toda Hamiltonian}
by using the symplectic Poisson bracket $J_{0}$, defined by $\left\{ q_{i},p_{j}\right\} =\delta_{ij}$.
In the $\left(a,b\right)$ variables, the symplectic Poisson bracket
$J_{0}$ becomes a quadratic Poisson bracket $\pi_{2}$:\begin{equation}
\left\{ a_{i},a_{i+1}\right\} =-a_{i}a_{i+1}\;,\quad\left\{ a_{i},b_{i}\right\} =a_{i}b_{i}\;,\quad\left\{ a_{i},b_{i+1}\right\} =-a_{i}b_{i+1}.\label{eq:Quadr Poisson bracket}\end{equation}
From this Poisson bracket and the Hamiltonian \eqref{eq:Linear Hamiltonian}
one obtains the equations of motion in the $\left(a,b\right)$ coordinates
\begin{equation}
\dot{a}_{j}=a_{j}\left(b_{j}-b_{j+1}+a_{j-1}-a_{j+1}\right)\quad;\quad\dot{b}_{j}=b_{j}\left(a_{j-1}-a_{j}\right).\label{eq:EOM Rel Toda lattice}\end{equation}

The Toda lattice is an integrable model. It also has a bi-Hamiltonian
structure, whose properties are summarized in \prettyref{sec:App-BiHamiltonian-structure}.
In order to construct this bi-Hamiltonian structure one needs to identify
two Hamiltonian functions $h_{1},h_{2}$ and two compatible Poisson
tensors $\pi_{1},\pi_{2}$ satisfying the same equations of motion,
i.e. if $\nabla=\left(\frac{\partial}{\partial x^{1}},...,\frac{\partial}{x^{M}}\right)$
where $x^{i}$ are our phase space coordinates, then \begin{equation}
\frac{d\overline{x}}{dt}=\pi_{1}\nabla h_{2}=\pi_{2}\nabla h_{1}\,.\label{eq:EOM-bihamiltonian-pair}\end{equation}

We already have the Hamiltonian function $h_{1}=\mbox{Tr}L$ \prettyref{eq:Linear Hamiltonian},
and the corresponding quadratic Poisson bracket $\pi_{2}$ \prettyref{eq:Quadr Poisson bracket}
such that $\pi_{2}\nabla h_{1}$ gives the equations of motion \prettyref{eq:EOM Rel Toda lattice}
\cite{Bruschi:1988,Oevel:1989}. A compatible linear Poisson bracket
$\pi_{1}$ was found in \cite{Oevel:1989pa} such that \[
\left\{ a_{i},b_{i}\right\} =a_{i}\;;\quad\left\{ a_{i},b_{i+1}\right\} =-a_{i}\;;\quad\left\{ b_{i},b_{i+1}\right\} =a_{i}\,,\]
 which together with the Hamiltonian $h_{2}=\frac{1}{2}\mbox{Tr}\left(L^{2}\right)$
also gives the equations of motion \prettyref{eq:EOM Rel Toda lattice}.
These two pairs make a bi-Hamiltonian system with equations of motion
given by  \eqref{eq:EOM-bihamiltonian-pair}. If we now construct
the master symmetries that obey the properties shown in \prettyref{sec:App-BiHamiltonian-structure}
\cite{Oevel:1989,Damianou:1994}, it becomes possible to construct
the hierarchy of Poisson brackets with the same equations of motion:%
\footnote{The Hamiltonian function $h_{0}$ is singular, reason why that point
is skipped from the sequence. $h_{0}$ can only be defined as a limit
\[
h_{0}=\lim_{\varepsilon\rightarrow0}\frac{1}{\varepsilon}\mbox{Tr}\left(L^{\varepsilon}\right)\sim\mbox{Tr}\left(\ln L\right).\]
 But to continue the sequence after this point, one can just do power
counting, followed by a direct verification of the equations of motion.%
}\[
\cdots=\pi_{0}\nabla h_{3}=\pi_{1}\nabla h_{2}=\pi_{2}\nabla h_{1}=\pi_{?}\nabla h_{-1}=\cdots\,.\]

Our final objective is to make an analogy with the system of $N$
magnons. Recalling the RS Lax matrix $\mathcal{L}_{rs}$, the sine-Gordon
model corresponds to the Hamiltonian $H\propto\mbox{Tr}\mathcal{L}_{rs}$
with the canonical Poisson brackets, while the system of magnons was
conjecture to correspond its {}``inverse\char`\"{} $H_{m}\propto\mbox{Tr}\mathcal{L}_{rs}^{-1}$,
with some other Poisson structure. In the limit we are considering
(relativistic Toda), we have\[
H_{sg}\propto\mbox{Tr}\mathcal{L}_{rs}\rightarrow h_{1}\quad;\quad H_{mag}\propto\mbox{Tr}\mathcal{L}_{rs}^{-1}\rightarrow h_{-1}.\]
 So, having started from the Hamiltonian $h_{1}=\mbox{Tr}L$, with
a quadratic Poisson bracket $\pi_{2}$, we want to find the Poisson
bracket corresponding to $h_{-1}=-\mbox{Tr}\left(L^{-1}\right)$ that
gives origin to the equations of motion \prettyref{eq:EOM Rel Toda lattice},
i.e. \[
\pi_{2}\nabla h_{1}=\pi_{m}\nabla h_{-1}.\]
To do so, we will restrict ourselves to $N=2$.

For the $N=2$ case, the Lax matrix reduces to \[
L=\left(\begin{array}{cc}
a_{1}+b_{1} & a_{1}\\
b_{2} & b_{2}\end{array}\right),\]
 the Hamiltonian functions are given by \begin{eqnarray*}
h_{1} & = & \mbox{Tr}L=a_{1}+b_{1}+b_{2};\\
h_{2} & =\frac{1}{2} & \mbox{Tr}L^{2}=\frac{1}{2}\left(a_{1}^{2}+2a_{1}b_{1}+2a_{1}b_{2}+b_{1}^{2}+b_{2}^{2}\right),\end{eqnarray*}
 and the corresponding Poisson bracket matrices are \[
\pi_{1}=\left(\begin{array}{ccc}
0 & a_{1} & -a_{1}\\
-a_{1} & 0 & a_{1}\\
a_{1} & -a_{1} & 0\end{array}\right)\quad;\quad\pi_{2}=\left(\begin{array}{ccc}
0 & a_{1}b_{1} & -a_{1}b_{2}\\
-a_{1}b_{1} & 0 & 0\\
a_{1}b_{2} & 0 & 0\end{array}\right).\]
 The EOM obtained from this bi-Hamiltonian system is\[
\left(\begin{array}{c}
\dot{a_{1}}\\
\dot{b_{1}}\\
\dot{b_{2}}\end{array}\right)=\pi_{1}\nabla h_{2}=\pi_{2}\nabla h_{1}=\left(\begin{array}{c}
a_{1}\left(b_{1}-b_{2}\right)\\
-a_{1}b_{1}\\
a_{1}b_{2}\end{array}\right).\]
 The objective is to determine which $\pi_{m}$ given origin to the
previous EOM with Hamiltonian function\[
h_{-1}=-\mbox{Tr}L^{-1}=-\frac{1}{b_{1}b_{2}}\left(a_{1}+b_{1}+b_{2}\right).\]
 We want to construct the next Poisson brackets given the master symmetries.
The master symmetries $X_{1}$ and $X_{2}$ were determined in \cite{Damianou:1994},
such that: $X_{1}\left(h_{n}\right)=\left(n+1\right)h_{n+1}$ and
$X_{2}\left(h_{n}\right)=\left(n+2\right)h_{n+2}$. These are given
by \begin{eqnarray*}
X_{1,2} & = & r_{1,2}^{1}\frac{\partial}{\partial a_{1}}+s_{1,2}^{1}\frac{\partial}{\partial b_{1}}+s_{1,2}^{2}\frac{\partial}{\partial b_{2}},\\
\mbox{with} &  & r_{1}^{1}=a_{1}^{2}+3a_{1}b_{2}\,;\; r_{2}^{1}=a_{1}\left(a_{1}^{2}+5a_{1}b_{1}+4b_{1}^{2}+2b_{1}b_{2}-b_{2}^{2}\right);\\
 &  & s_{1}^{1}=b_{1}^{2}+2a_{1}b_{1}\,;\; s_{2}^{1}=b_{1}\left(-2a_{1}^{2}-a_{1}b_{1}-2a_{1}b_{2}+b_{1}^{2}\right);\\
 &  & s_{1}^{2}=b_{2}^{2}-a_{1}b_{2}\quad;\; s_{2}^{2}=b_{2}\left(2a_{1}^{2}+3a_{1}b_{1}+4a_{1}b_{2}+b_{2}^{2}\right).\end{eqnarray*}
 Then the next Poisson Brackets are given by property 4 (recall that
the Poisson matrices are anti-symmetric):%
\footnote{To determine the Lie derivative of the 2-tensor $\pi^{ij}$, we use
the rule for a general tensor\begin{eqnarray*}
\mathcal{L}_{X}T_{\qquad b_{1}...b_{s}}^{a_{1}...a_{r}} & = & X^{c}\nabla_{c}T_{\qquad b_{1}...b_{s}}^{a_{1}...a_{r}}-\nabla_{c}X^{a_{1}}T_{\qquad b_{1}...b_{s}}^{c...a_{r}}-\cdots-\nabla_{c}X^{a_{r}}T_{\qquad b_{1}...b_{s}}^{a_{1}...c}+\\
 &  & \qquad+\nabla_{b_{1}}X^{c}T_{\qquad c...b_{s}}^{a_{1}...a_{r}}+\cdots+\nabla_{b_{s}}X^{c}T_{\qquad b_{1}...c}^{a_{1}...a_{r}}.\end{eqnarray*}
}\begin{eqnarray*}
\pi_{3} & = & -\mathcal{L}_{X_{1}}\pi_{2}=\left(\begin{array}{ccc}
0 & a_{1}b_{1}(a_{1}+b_{1}) & -a_{1}b_{2}(a_{1}+b_{2})\\
 & 0 & -a_{1}b_{1}b_{2}\\
 &  & 0\end{array}\right);\\
\pi_{4} & = & -\frac{1}{2}\mathcal{L}_{X_{2}}\pi_{2}=\left(\begin{array}{ccc}
0 & a_{1}b_{1}((a_{1}+b_{1})^{2}+a_{1}b_{2}) & -a_{1}b_{2}(a_{1}(a_{1}+b_{1})+2a_{1}b_{2}+b_{2}^{2})\\
 & 0 & -a_{1}b_{1}b_{2}(a_{1}+b_{1}+b_{2})\\
 &  & 0\end{array}\right).\end{eqnarray*}
 With these results we can easily see that $\pi_{3}\nabla h_{-1}$
does not give the right equations of motion, but $\pi_{4}\nabla h_{-1}$
does. So the Hamiltonian $h_{-1}$ with Poisson bracket $\pi_{4}$
will give the same classical behavior than the Hamiltonian $h_{1}$
with Poisson bracket $\pi_{2}$. The hierarchy is given by (the point
$\pi_{3},h_{0}$ is not defined) \[
\cdots=\pi_{0}\nabla h_{3}=\pi_{1}\nabla h_{2}=\pi_{2}\nabla h_{1}=\pi_{4}\nabla h_{-1}=\cdots\,.\]

All of these pairs generate the same equations of motion
and time delay. In particular, the Hamiltonian $h_{-1}$ with (quartic)
Poisson bracket $\pi_{4}$ will give the same classical behavior than
the Hamiltonian $h_{1}$ with Poisson bracket $\pi_{2}$. Since we
have found that ( in the limit of well separated magnons)  the Hamiltonian
reduces to \[
H_{mag}=\mbox{Tr}\mathcal{L}_{rs}^{-1}\rightarrow h_{-1},\]
it will reproduce the correct equations of motion (the same as the limiting
case of  sine-Gordon solitons) as long as we use the quartic
Poisson structure $\pi_{4}$ defined above.

For a non degenerate Poisson structure, the phase shift is given by
the corresponding symplectic form (the inverse of the Poisson tensor).
The usual symplectic form is replaced with the following\[
\int p_{i}\dot{q}_{i}\rightarrow\int p_{i}\left(\pi^{-1}\right)_{ij}\dot{q_{j}}.\]
 (For a degenerate Poisson structure one has to check this more carefully.)
Consequently, if two different systems have the same equations of
motion, the different Poisson Structures give origin to different
phase shift.

\section{Conclusions \& Acknowledgments\label{sec:Conclusions-&-Acknowledgments}}
We have in the present paper considered the question of an N-particle dynamics
that would fully describe interacting magnons  at the semiclassical level. For this we have
specified the interacting hamiltonian as a member of the RS hierarchy. This hamiltonian
had the property that it reproduces energies of magnons. We argued that an alternative symplectic
form is needed in order to obtain the correct magnon phase shifts. We have considered the
question of the modified symplectic form explicitly for the case of well separated magnons.
In this limit one had the results of relativistic Toda theory where a sequence of symplectic
forms was already established in the literature.

Altogether the new hamiltonian and the modified symplectic form are defined so to
reproduce correctly the original classical equations of motion and therefore the time delay.
Regarding future interesting problems we mention the following. We have succeeded in establishing
the necessary symplectic form in the limit of well separated magnons. For establishing an
exact result one will have to give the multi-Poisson structure for the RS model itself. It is
likely that this is definitely possible, although technically (and possibly conceptually) challenging.
But one can definitely expect that a sequence of symplectic structures always follows for an
integrable system. Generalization of the present construction to magnons moving on
higher spheres \cite{Chen:2006gea} is also a challenging task. One would also want to define the dynamics in the periodic case appropriate for string motions with finite J \cite{Klose:2008rx}.

One of us (AJ) would like to acknowledge the pleasant hospitality of L. Feher  and
and the organizers of the Oetvos Summer School in Budapest where part of this work was done.
We also acknowledge discussions with Kewang Jin, Anastasia Volovich  and Mark Spradlin
on related topics, We are grateful to Jean Avan for his reading  of the manuscript and for his
constructive suggestions. This work was supported in part by DOE grant DE-FG02-91ER40688- Task
A. I. Aniceto was also supported in part by POCI 2010 and FSE, Portugal,
through the fellowship SFRH/BD/14351/2003.

\appendix

\section{Bi-Hamiltonian structure of Relativistic Toda Lattice \cite{Bruschi:1988,Oevel:1989,Suris:1993,Damianou:1994}
\label{sec:App-BiHamiltonian-structure}}

The relativistic Toda lattice is an integrable model, and has a sequence
of conserved quantities \[
h_{n}=\frac{1}{n}\mbox{Tr}\left(L^{n}\right).\]
 To this sequence we have a corresponding set of Hamiltonian vector
fields $\chi_{1},...,\chi_{n}$, where $\chi_{i}=\left[\pi,h_{i}\right]$,
where $\pi$ is some Poisson structure and $\left[,\right]$ is the
Schouten bracket (Lie bracket). Also, we have a hierarchy of Poisson
2-tensors $\pi_{1},...,\pi_{n}$ (which are polynomial homogeneous
of degree $n$), and a sequence of master symmetries $X_{1},...,X_{n}$,
which obey the following properties (more information on the properties
of these entities can be found in \cite{Damianou:1994} and references
therein):

\begin{enumerate}
\item the $\pi_{n}$ tensors are all Poisson structures. The corresponding
Poisson brackets are given by \[
\left\{ f,g\right\} =\sum_{i,j}\pi^{ij}\frac{\partial f}{\partial x^{i}}\wedge\frac{\partial g}{\partial x^{j}},\; f,g\in C^{\infty}.\]
 Note that $\pi^{ij}$ are the matrix elements of the matrix $\pi_{n}$
corresponding to this 2-tensor, and $\overline{x}=\left(x^{1},...,x^{M}\right)$
are the coordinates of the Hilbert space, in our case $\left(a_{1},...,a_{N-1},b_{1},...,b_{N}\right)$;
\item functions $h_{n}$ are in involution with all $\pi_{m}$;
\item $X_{n}\left(h_{m}\right)=\left(n+m\right)h_{m+n}$;
\item $\mathcal{L}_{X_{n}}\left(\pi_{m}\right)\equiv\left[X_{n},\pi_{m}\right]=\left(m-n-2\right)\pi_{n+m}$,
where $\mathcal{L}_{X}$ in the Lie derivative in the direction of
the vector$X$;
\item $\left[X_{n},X_{m}\right]=\left(m-n\right)X_{n+m}$;
\item \begin{equation}
\pi_{n}\nabla h_{m}=\pi_{n-1}\nabla h_{m+1},\label{eq:Hierarchy of EOM}\end{equation}
 where $\pi_{n}$ now denotes the Poisson matrix of the tensor $\pi_{n}$,
and $\nabla=\left(\frac{\partial}{\partial x^{1}},...,\frac{\partial}{x^{M}}\right)$.
\end{enumerate}
It is known \cite{Magri:1978} that once our system is bi-Hamiltonian,
which means that we can identify two Hamiltonian functions $h_{1},h_{2}$
and two compatible Poisson tensors $\pi_{1},\pi_{2}$ satisfying \[
\pi_{1}\nabla h_{2}=\pi_{2}\nabla h_{1},\]
 then we can find the whole hierarchy stated above, and the equations
of motion are just given by\begin{equation}
\frac{d\overline{x}}{dt}=\pi_{1}\nabla h_{2}=\pi_{2}\nabla h_{1}=\pi_{0}\nabla h_{3}=\cdots\,.\end{equation}
 All of these properties are valid for $m,n>0$ but can be seen to
generalize to negative values as well.

In the case one of the Poisson brackets is symplectic, then one can
find a recursion operator which can then be applied to the initial
symplectic bracket to determine the hierarchy \cite{Das:1988wm,Oevel:1989}.
In our case, we will see that in the $\left(a,b\right)$ coordinates
the Poisson brackets are not symplectic (not even non degenerate,
as we don't have the same number of $a's$ and $b's$) and it is non-trivial
to find an extra Poisson bracket in the $\left(p,q\right)$ coordinates
apart from the symplectic one, in order to form a bi-Hamiltonian system
\cite{Suris:1993}. In this case we will turn to the problem of finding
master symmetries that allow us to construct the hierarchy of Poisson
structures.

\bibliographystyle{D:/Home/nes/Trabalho/Work/Bib_styles/utcaps}
\bibliography{relcal}

\end{document}